\documentstyle[12pt]{article}
\begin{document}
\renewcommand{\thefootnote}{\fnsymbol{footnote}}
\def\thefootnote{\arabic{footnote}}
\setcounter{footnote}{1}
\def\thesection{\Roman{section}.}
\parindent 0mm
\topskip 20mm

\vspace{20mm}

\begin{center}
\large
{\bf
Effect of the Haar measure on the finite temperature
effective potential of $SU(2)$ Yang-Mills theory

}

\vspace{5mm}

\normalsize
\baselineskip 10mm
K. Sailer$^{a}$,
  A. Sch\"afer$^{b}$, W. Greiner$^{b}$

\vspace{5mm}

$^a$ {\em  Department for Theoretical Physics,
          Kossuth Lajos University, H-4010 Debrecen, Hungary.}\\
$^b$ {\em Institut f\"ur Theoretische Physik,
Johann Wolfgang  Goethe Universit\"at,
D-60054 Frankfurt am Main, Germany  }\\

\end{center}

\vspace{8mm}

\begin{abstract}
\normalsize
\rm
\parindent 10mm
\baselineskip  10mm

Including the Haar measure
we show that the effective potential of the  regularized
 SU(2) Yang-Mills theory
 has a minimum at vanishing Wilson-line  $W=0$ for strong coupling,
 whereas it develops
two degenerate minima close to $W=\pm 1$ for weak coupling.
This suggests that the non-abelian character of $SU(2)$ as
 contained
in the Haar measure might be responsible for confinement.

PACS number(s): 11.15.Tk, 11.10.Lm
\end{abstract}

\newpage
\normalsize
\rm
\textheight 21.0cm
\headheight -1. cm
\oddsidemargin 1mm
\leftmargin 1mm
\topmargin -5mm
\textwidth 15.5cm
\parindent 10mm
 \baselineskip 10.0mm
\topskip 0mm
\footheight 1.5cm
\footskip 5mm

Concerning
the finite temperature effective potential
of $SU(2)$ Yang-Mills theory
it was established by Weiss \cite{Wei81}
 that the tree level contribution of the timelike
`gluons' due to the Haar measure is cancelled by a piece of the
1-loop contribution of the longitudinal gluons. This finding has a
far reaching importance for the understanding of confinement. If
the
non-abelian character of $SU(2)$ is the crucial factor for
confinement one would expect the Haar measure to play an important
role. Consequently, it should induce a contribution to the effective
potential dominating at large distances. On the other hand if it
does
not contribute at all (as suggested by the results of Weiss) this
supports the idea propagated e.g. by Gribov \cite{Gri93} that only
the strength of the interaction is crucial.

The 2--loop contributions of the order $g^2$
to the effective
potential have been established in the perturbative
regime \cite{Kap79}, i.e. for non-vanishing Wilson line $W=\pm 1$.
Allowing for arbitrary values of $W$,
we show now explicitly  that the cancellation   observed by Weiss
does not hold in the order $g^2$, i.e.
 the 1-loop contribution
of the Haar measure
is not cancelled by the 2-loop contribution due to the
 usual Yang-Mills self-interaction.
Thus the Haar measure contributes to the effective potential
and leads to a
 minimum at
$W=0$ for sufficiently strong coupling.
A similar result has been obtained for $SU(2)$ lattice gauge theory
in Ref. \cite{Pol82}.
For weak coupling we shall recover the results obtained in Ref.
 \cite{Wei81}.

General arguments were given that a deconfining phase transition
occurs with increasing temperature if the Yang-Mills theory is
 confining
at zero temperature \cite{Pol78,Sus79}.
It was also shown  that lattice gauge theory does not confine
 static
quarks if the Haar measure
 is replaced by the Euclidean
one \cite{Pat81}. Here we show that
the minimum  found at $W=0$  corresponds to confinement.
 How the continuum
limit  can be taken and to what extent the non-trivial minimum does
 survive  remains of course an open question which can only
be answered by  renormalization group analysis being out of the
scope  of the present letter.

Similarly to \cite{Wei81} we use
 a time independent, diagonal gauge,
and periodic boundary conditions for the spatial components of the
vector potential
 in the `time' direction.
We allow for the non-vanishing vacuum expectation value
 $v \equiv \langle a^{-1} \beta g A^{03}
\rangle \equiv a^{-1} \beta C $
 ($A^{\mu a}$  the gluon
 vector potential, $\mu$ and $a$ the
 Lorentz  and $SU(2)$ colour indices, respectively).
Throughout this paper we use the notations of
 Ref. \cite{Wei81} and use the same cut-off regularization. The UV
cut-off $\Lambda$ is interpreted in terms of the lattice spacing $a$
used for the definition of the path integral via
$a^{-3} = (2\pi )^{-3} \int_0^\Lambda d^3 k $ with $\Lambda =
(6\pi^2 )^{1/3} a^{-1}  $ and $d^3 k$ the volume element in
3-momentum space. The same spacing $a$ is assumed in `time'
direction.

Let us write for the timelike gluon field $A^{03} ({\vec x} )
=  C/g + \delta \phi ({\vec x})$ and for the spatial components
$A^{ia } ( {\vec x}, t)$ $(i=1,2,3$; $a=1,2,3)$.
Expanding the Haar measure  in powers of the
 fluctuation $\delta \phi ({\vec x})$ and including all terms up to
the order $g^2$,
we obtain the  tree level effective action,
$   S_{eff} =  S_0 + S_1 + S_2   $
where
\begin{eqnarray}
    S_0 &=&
 \frac{1}{2a^4} \int_0^\beta dt \int d^3 x
  \left\lbrack a^2 ( \nabla \delta \phi )^2
            + (a \partial^0  A^i + C {\hat 3} \times  A^i
              )^2
   + \frac{1}{2} a^2 ( \partial^i  A^j - \partial^j  A^i )^2
             \right\rbrack
            \nonumber\\
   & &  - \frac{1}{a^3 }
      \int  d^3 x  \ln (1 - \cos v)
            + \frac{1}{2a^3} \frac{ g^2 a^{-2} \beta^2 }{
         1 - \cos v}
           \int d^3 x ( \delta \phi )^2
\end{eqnarray}
 (${\hat 3}$ is the unit vector in the
direction 3 of  colour space),  $S_1$ and $S_2$ are
 the cubic and quartic self-interaction terms,
respectively.
 The last term of
$S_0$ proportional to
$g^2$ was neglected in Ref. \cite{Wei81}, as well as the
self-interaction  $S_1$ and $S_2$.
In order to
calculate the finite temperature effective potential, we determine
the partition function $Z$  treating  $S_1$ and $S_2$
as perturbations of the `free' theory
defined by  $S_0$.
The free propagator of the field $\delta \phi$ acquires now the
mass term
$ a^{-1} g^2 \beta ( 1 - \cos v )^{-1} \equiv  M^2 a^2 $
 due to the Haar measure generated quadratic
 self-interaction.

The effective potential is given in terms of
the one-particle irreducible part of the logarithm
of the partition function,
$V_{eff} (\beta ; v)
= - (\beta V)^{-1} \ln Z_{\mbox{1PI}}
= V_0 (\beta ; v) + \Delta V (\beta ; v)$.
The effective potential for the
`free' theory
$V_0 (\beta ; v) = V_W (\beta ; v) + V_H (\beta ; v)$
consists of the term $V_W (\beta ; v)$ of the order $g^0$
 found by Weiss \cite{Wei81} and the term  $V_H ( \beta ; v )$   of
the order $g^2$
generated by the Haar measure:
\begin{eqnarray}
   V_H ( \beta ; v) = \frac{ 1}{ \beta V}
   \int \frac{ d^3 k}{ (2\pi )^3 }
        \ln \left( \frac{ g^2 a^{-1} \beta}{
  a^2 k^2 ( 1 - \cos v) }  + 1  \right)
       \approx  \frac{ g^2 }{ a^4 \alpha_0 \sin^2 (v/2)  }
                + {\cal O} (g^4 )
\end{eqnarray}
with $\alpha_0 = (6\pi^2 )^{1/2}$ for $\beta /a \to \infty$.
The 2-loop contribution  $\Delta V (\beta ; v)$
 was obtained by carrying out the Matsubara sums with the standard
techniques of finite temperature field theory \cite{Kapbook}.
 UV-divergences have been
removed by subtracting from each 2-loop diagram  those with the same
structure but a single loop taken in the limit $\beta \to \infty$ in
all possible ways.
The IR momentum cut-off $\mu$  was chosen in an UV cut-off $(a)$
 dependent way by requiring that the
free energy of the perturbative vacuum does not depend on $\mu$.
 Then the
 free energy of the perturbative vacuum
turned out to be a temperature independent constant which
 was subtracted. As a result
of this choice of the IR cut-off $\mu $ all terms of $\Delta V
 (\beta , v)$
 depending on  $\mu$ vanish in the $\beta /a \to \infty$ limit
and we obtain:
\begin{eqnarray}
\label{DeltaV}
      \Delta V (\beta , v) &=&
    \frac{g^2}{24 \beta^4}
    \left\{  1  +
       \frac{8}{\pi^2} \left| \sin \frac{v}{2} \right|
     \left(  \left| \sin \frac{v}{2}  \right| - \frac{2\pi }{3}
     \right)
   + \frac{1}{\pi^4} \left( \frac{ 4\pi^2}{3} + G(v)
                               \right)   G(v)      \right.
                   \nonumber\\
     &   &         \left.
 + \frac{21}{4\pi^4 e^2}  \frac{\sin^2 v}{ (\cosh 1 - \cos v
        )^2  }  ( 1 - e^{-1} )( 2 - 5e^{-1} )
            \right\}
\end{eqnarray}
with
\begin{eqnarray}
    G(v) &=& \frac{2}{e} \left(  \frac{ \cos v - e^{-1} }{
      1 + e^{-2} - 2 e^{-1} \cos v  }
    -   \frac{ 1}{ 1 - e^{-1}  }    \right)
\end{eqnarray}
and $e$ the basis of the natural logarithm.
Our conclusion is that the 2-loop contribution $\Delta V$ due to
the non-abelian gluon self-interaction $S_1 + S_2$  does not cancel
 the
1-loop contribution $V_H$ due to the Haar-measure.
We also see that the $v$-dependent terms of
$\Delta V $ vanish for vanishing background field $v=0$.  For $v=0$
we obtained $\Delta V_{v=0} = \frac{1}{24} g^2 \beta^{-4}$ which is
identical to the 2-loop contribution of gluons and ghosts found in
\cite{Kap79}. It needs further investigation how our results would
be
modified by using lattice regularized propagators for the
 calculation of
the loop integrals.

Let us express the effective potential in terms of the
vacuum expectation value
of the Wilson line operator,
\begin{eqnarray}
   W ( {\vec x} )
 &=& \left\langle
 \frac{1}{2} {\mbox{tr}}
  \left( \prod_{j =1}^{\beta /a}
   \exp \left\{ {\rm i} g \phi ( {\vec x} ) \tau^3 /2 \right\}
  \right)      \right\rangle
        \approx \cos \left( \frac{v}{2} \right)
 \equiv W
\end{eqnarray}
for $R = g^2 (\beta /a)^4$ fixed and $\beta /a \to \infty$
 $(| W | \le 1)$.
For $\beta /a \to \infty$ we can neglect the 2-loop
contribution $\Delta V$ as compared to the Haar measure term $V_H$.
It turns out that $v=0$, i.e. $W=0$ is
 a minimum of the effective potential
   for $R > R_c \equiv 2\alpha_0 /3$,
 i.e. for
strong coupling and it turns over in a maximum for $R < R_c$, i.e.
for weak coupling. In the latter case we obtain two minima
positioned at $W \approx \pm ( 1 -  (R/8R_c )^{1/2}  )$ for $R\ll
R_c$, and at $W \approx \pm ( 1 - (R/R_c) )^{1/2}$ for $R \approx
R_c$ $(R<R_c )$. Thus the effective potential in the approximation
used exhibits the features of a second order phase transition with
the order parameter $| W |$ vanishing smoothly for $ R \to R_c$
(see Fig. 1).
The infinite values of $V_H$ at $W
= \pm 1$ occur due to the fact that
 the expansion used for the Haar-measure potential
 is not valid for $v = 0$ and $\pm 2\pi$.
Unluckily
the critical temperature $T_c$ defined by $g^2 (aT_c)^{-4} = R_c
=5.1$
decreases rather slowly with decreasing coupling
($aT_c \sim 0.94$ to $0.71$ for $4/g^2 \sim 1$ to $ 3$) and
does not agree quantitatively
with the numerical results of lattice gauge simulations
\cite{Pol82}. The relation between these two approaches has to be
further investigated.

It is rather intriguing to extract information on the behaviour of
the correlator of Wilson-lines.
For strong coupling the vacuum is characterized  by  $v=\pi $ and
 the
free energy of a static  quark-antiquark pair increases linearly
with their separation distance in leading order:
\begin{eqnarray}
   F_{ q {\bar q} } &=& - \beta^{-1} \ln \langle
     {\cal W} ( {\vec x} ) {\cal W} ( {\vec y} )  \rangle
     = \beta^{-1} M | {\vec x} - {\vec y} |
      + \beta^{-1} \ln \frac{ | {\vec x} - {\vec y} | }{a}
      + {\mbox{ const. }}
\end{eqnarray}
This corresponds to the string tension
 $   \kappa = \beta^{-1} M
  = \beta^{-2} \lbrack aR/(2\beta ) \rbrack^{1/2}  $.
For weak coupling $R \ll 1$ the degenerate vacua are at $v=\epsilon$
 and
$\pm ( 2\pi - \epsilon )$ with $0 < \epsilon^2 =
( 8R/R_c )^{1/2} \ll 1$ and  the correlator of the
Wilson lines  the
 Debye-screened Coulomb law $  F_{ q {\bar q} } \sim
e^{ - M | {\vec x} - {\vec y} | } / | {\vec x} - {\vec y} |$
with the screening length $1/M
  =  \beta (\beta /a)^{1/2}
    \sqrt{ 2/R}  \to \infty $
for $R \to 0$.

Summarizing, we established that the
finite temperature  effective potential
of the  regularized $SU(2)$ Yang-Mills theory
exhibits rather different
qualitative behaviour in the strong coupling and weak coupling
 limits.
For strong coupling $(R \gg R_c )$ the effective potential has
 a single
minimum for
vanishing Wilson-line $W=0$ $(v=\pi )$, and the vacuum
state is confining. On the other hand, there are
degenerate minima at $W \to 1$ $(v \to 0)$ and $W \to -1$
 $(v \to \pm 2\pi )$
and a maximum at $W=0$ for weak coupling $(R \to 0)$. These minima
correspond to a vacuum state in which static quarks are not
 confined.

The results obtained  hint to the
possibility that the Haar measure induced vertices could be
 responsible
for confinement. If proven this would be of fundamental importance.
For a proof, however, a resummation of the contributions of all
induced vertices to
the effective potential is needed. Furthermore an investigation of
 the
renormalization group trajectories is needed to establish
the effective potential in the continuum limit.

{\bf Acknowledgement.} One of the authors (K.S.) is greatly
indebted to
J. Polonyi for the valuable discussions.

\newpage

{\bf Figure Caption}

{\bf Fig. 1} The order parameter $|W|$ of the deconfining phase
 transition vs.
 $R/R_c$.


\begin{thebibliography}{99}
\bibitem{Wei81}
N. Weiss, Phys. Rev. {\bf D24}, 475 (1981);
        Phys. Rev. {\bf D25}, 2667 (1982).
\bibitem{Gri93}
V.N. Gribov, {\em  Orsay Lectures on Confinement I and II},
(June 1993), hep-ph-9403218, hep-ph-9404332.
\bibitem{Kap79}
J. I. Kapusta, Nucl. Phys. {\bf B148}, 461 (1979).
\bibitem{Pol82}
J. Pol\'onyi, K. Szlach\'anyi, Phys. Lett. {\bf 110B}, 395 (1982).
\bibitem{Pol78}
A. M. Polyakov, Phys. Lett. {\bf 72B}, 477 (1978).
\bibitem{Sus79}
L. Susskind, Phys. Rev. {\bf D20}, 2610 (1979).
\bibitem{Pat81}
A. Patrascioiu, E. Seiler, I. O. Stamatescu,
Phys. Lett. {\bf B107}, 364 (1981).
\bibitem{Polo91}
J. Pol\'onyi,  in {\em Quark-Gluon Plasma}, ed. by R.C. Hwa
(World Scientific, Singapore, 1990), p. 1.
\bibitem{Joh91}
K. Johnson, L. Lellouch, and J. Pol\'onyi,  Nucl. Phys.
{\bf B367}, 675 (1991).
\bibitem{Kapbook}
J.I. Kapusta, {\em Finite Temperature Field Theory},
(Cambridge, University Press, 1989).
\end{thebibliography}
\end{document}